\documentclass[preprint2]{aastex}
\usepackage{graphicx}

\begin{document}

\title{
Application of Monte Carlo-based statistical significance determinations to the Beta Cephei stars
\\V400 Car, V401 Car, V403 Car and V405 Car}

\author{C. A. Engelbrecht}

\affil{Department of Physics, University of Johannesburg,
P. O. Box 524, Auckland Park 2006, South Africa}\email{chrise@uj.ac.za}

\author{F.A.M. Frescura}

\affil{Centre for Theoretical Physics, University of the Witwatersrand,
Private Bag 3, WITS 2050, South Africa}\email{Fabio.Frescura@wits.ac.za}

\author{B. S. Frank}

\affil{School of Physics, University of the Witwatersrand, Private Bag 3, WITS 2050, South Africa}

\begin{abstract}

We have used Lomb-Scargle periodogram analysis and Monte Carlo significance tests to detect
periodicities above the 3-sigma level in the Beta Cephei stars V400 Car, V401 Car, V403 Car and
V405 Car. These methods produce six previously unreported periodicities in the expected frequency
range of excited pulsations: one in V400 Car, three in V401 Car, one in V403 Car and one in V405
Car. One of these six frequencies is significant above the 4-sigma level. We provide statistical
significances for all of the periodicities found in these four stars.\\ \ \\

\end{abstract}

\section{Introduction}

The refinement of asteroseismological models requires increased numbers of statistically reliable
pulsational periodicities for observed stars. Recent advances in the detailed modelling of Beta
Cephei stars (for example, \citet{miglio07}, \citet{pamyatnykh07}, \citet{ smolec07}) have added
urgency to the need for accurate and complete identification of pulsation modes in {\em real}
stars. This requires quantified significances for peaks identified in periodograms.
\citet{frescura08} recently reviewed some statistical significance tests that are in common use in
periodogram analysis, and have assessed and compared their reliability. In their paper, they
recommend implementing Monte Carlo methods to establish the significances of periodogram peaks,
arguing that these are more reliable than the methods in common use. Accordingly, we have
calculated the Lomb-Scargle periodograms, suitably normalised by the variance of the data
\citep[as advised by][]{hb86}, for observations of the four Beta Cephei stars V400 Car, V401 Car,
V403 Car and V405 Car, and have applied Monte Carlo methods in the way proposed by
\citet{frescura08} to assign significances to peaks identified in each periodogram analysis. This
involved the construction of cumulative distribution functions (CDF's) for pseudo-Gaussian noise
generated on the observation time-grid of each of the respective real data sets. These CDF's were
then used to determine quantitatively the significances of candidate peaks in their respective
periodograms. The results of these analyses are described in the sections that follow.

We report a number of newly discovered periodicities that lie above the 3 sigma level (a
significance of 99.7\%) in each of these stars. Our analysis determines that there are seven
periodicities above the 3-sigma level that lie within the expected range for pulsation modes in
V400 Car, eight in V401 Car, four in V403 Car, and four in V405 Car. The methods applied in this
paper provide an objective assessment of the identified periodicities.

A canonical value of 4:1 for the signal-to-noise ratio in a Fourier amplitude
spectrum \citep[see][]{breger93} has long been applied as a threshold for
accepting that a detected periodicity may be attributed to a variation in the
stellar magnitude itself. \citet{kuschnig97} showed that this 4:1 ratio
corresponds to a significance of approximately 99.9\% as determined from Monte
Carlo trials. The detections claimed in this paper therefore correspond very closely to the
generally accepted threshold for period detection in pulsating stars.

\section{Significances of periodicities in V400 Car}

\begin{figure}
\epsscale{.80}
\plotone{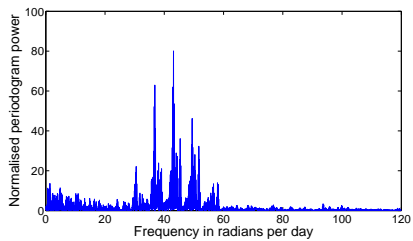}
 \caption{
Over-sampled Lomb-Scargle periodogram for V400 Car.}
 \label{figure1}
\end{figure}

\begin{figure}
\epsscale{.80} \plotone{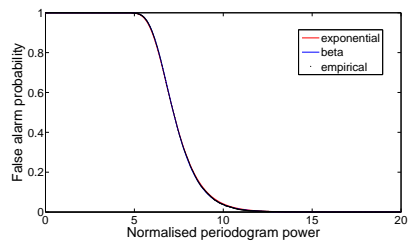}
 \caption{
Limiting empirical CDF for the V400 Car data set. Both exponential and
$\beta$-functions give excellent fits to the empirical CDF. Their deviations
from the empirical CDF can only be seen under magnification.}
 \label{figure2}
\end{figure}

The data set consists of 533 observations of V400 Car (star no. 11 in the
designation of \citet{feast58} and \citet{turner80} for stars in NGC3293) in
the Johnson B band, obtained with the 1.0 m telescope at the Sutherland
station of the South African Astronomical Observatory (SAAO) during a total
timespan of 68 days in 1984. The Lomb-Scargle periodogram of these data,
normalised by the variance of the data, is shown in Figure \ref{figure1}.
The periodogram was over-sampled by a factor of approximately 25. The limiting
empirical CDF computed for the time data, as described in \citet{frescura08},
is shown in Figure \ref{figure2}.

\begin{figure}
 \epsscale{.80}
 \plotone{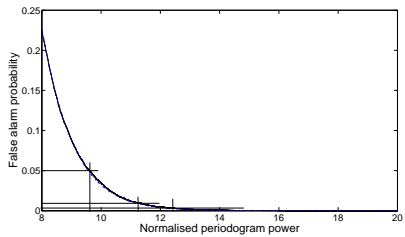}
 \caption{
Enlargement of the critical region of the CDF for V400 Car.}
 \label{figure3}
\end{figure}

An enlargement of the critical region of this CDF is shown in Figure
\ref{figure3}.
Two theoretical distribution functions were re-fitted to the tail of the empirical CDF shown in
the enlargement. These are the exponential distribution function and the incomplete beta function
(called simply the beta function in the remainder of this paper). A detailed discussion of this
refitting procedure and of these theoretical distribution functions is given in
\citet{frescura08}. Both theoretical distributions fit the empirical CDF's reasonably well in this
region, though not to the same degree. For V400 Car and for V401 Car, the peak significances
estimated from the empirical CDF's were slightly lower (i.e. the false alarm probabilities were
higher) than those obtained from the best-fitting theoretical distribution functions. For V403 Car
and for V405 Car, the peak significances estimated from the empirical CDF's lie mostly between
those obtained from the best-fits of the two respective theoretical distribution functions.
Comparisons of the respective estimates of significance obtained from the empirical CDF's and from
the theoretical distribution functions are displayed in Tables \ref{table1}, \ref{table3},
\ref{table5}, and \ref{table7}. For all four stars, the peak significances obtained from the
best-fitting exponential distribution function are lower than those obtained from the beta
function. In each of these four cases, therefore, the best-fitting exponential function provides
the most conservative estimate of significance for the detected periodicities, and this is the
estimate that we used in the tables that follow. In Figure \ref{figure3}, the horizontal solid
lines indicate false alarm probabilities of $0.3\%$, $1\%$ and $5\%$ respectively (i.e. peak
significances of $99.7\%$, $99\%$ and $95\%$ respectively). The periodogram peak values (i.e.
normalised power levels) associated with these significances are read off the abscissae where the
best-fit exponential function crosses the respective horizontal solid lines. The power levels
corresponding to the aforementioned three significance thresholds, for the empirical CDF as well
as for both of the theoretical distribution functions, are displayed in Table \ref{table1} .
\begin{table*}
 \centering
 \begin{minipage}{140mm}
\caption{Significance Levels for V400 Car} \vspace*{4mm}
      \begin{tabular}{@{}cccc@{}}
    \hline
 Significance & \multicolumn{3}{c}{Periodogram Power Level}  \\
 & empirical & exponential function & beta function \\
   \hline
 95 \% & 9.61 & 9.60 & 9.56\\
 99 \% & 11.23 & 11.23 & 11.13\\
 99.7 \% & 12.48 & 12.43 & 12.28\\
 \hline
\end{tabular}
\label{table1}
\end{minipage}
\end{table*}

The significances of peaks appearing in the periodogram of V400 Car
were determined by using the exponential function providing the best
fit to the tail of the associated empirical CDF. The oversampled,
normalised Lomb-Scargle periodogram shown in Figure \ref{figure1}
was subjected to a standard prewhitening procedure, as follows:
\begin{enumerate}
  \item
determine the frequency at which the highest peak occurs in the periodogram;
  \item
determine the best-fitting amplitude and phase of a sinusoid with this
frequency by a least-squares comparison with the data;
  \item
subtract this sinusoid (including the best-fitting amplitude and phase) from
the data;
  \item
recalculate the oversampled, normalised Lomb-Scargle periodogram for the
modified data;
  \item
repeat step (1) above;
  \item
determine the respective best-fitting amplitudes and phases of two sinusoids with the frequencies
determined in steps (1) and (5) above;
  \item
iterate steps (4) to (6) above, recalculating the respective best-fitting amplitudes and phases
for the entire set of determined frequencies each time.
  \end{enumerate}
Prewhitening was continued until a level clearly below the 99.7\% (3-sigma)
significance level (as established by the best-fit exponential function) of
the data set was reached. A full list of the results of the prewhitening
procedure for V400 Car appears in Table \ref{table2}.
\begin{table*}
 \centering
 \begin{minipage}{160mm}
\caption{Frequencies determined for V400 Car} \vspace*{4mm}
      \begin{tabular}{@{}cccccc@{}}
    \hline
 Number & Rad $d^{-1}$ & Cycles $d^{-1}$ & Amp (mmag) & Power & Significance \\
   \hline
 $f_1$ & 43.11 & 6.86 & 6.4 & 80.1 & $>99.9999$ \% \\
 $f_2$ & 42.24 & 6.72 & 4.1 & 50.1 & $>99.9999$ \% \\
 $f_3$ & 45.38 & 7.22 & 3.0 & 39.3 & $>99.9999$ \% \\
 $f_4$ & 44.36 & 7.06 & 2.9 & 42.1 & $>99.9999$ \% \\
 $f_5$ & 41.75 & 6.64 & 2.5 & 26.7 & $>99.9999$ \% \\
 $f_6$ & 2.36 & 0.38 & 2.4 & 29.2 & $>99.9999$ \% \\
 $f_7$ & 1.40 & 0.22 & 2.1 & 27.2 & $>99.9999$ \% \\
 $f_8$ & 3.29 & 0.52 & 1.5 & 18.5 & $99.9993$ \% \\
 $f_9$ & 0.66 & 0.11 & 1.3 & 17.8 & $99.9986$ \% \\
 $f_{10}$ & 41.40 & 6.59 & 1.2 & 12.9 & $99.81$ \% \\
 $f_{11}$ & 12.24 & 1.95 & 1.2 & 13.0 & $99.83$ \% \\
 $f_{12}$ & 53.48 & 8.51 & 1.1 & 12.8 & $99.79$ \% \\
 $f_{13}$ & 15.05 & 2.40 & 1.1 & 10.0 & $96.6$ \% \\
 \hline
\end{tabular}
\label{table2}
\end{minipage}
\end{table*}
For convenience, each frequency is listed in units of radians per day and of
cycles per day, along with its best-fit amplitude in millimagnitudes, its
respective normalised power in the periodogram, and the associated
significance of the normalised power level, as determined from the fitting of
the exponential distribution function to the tail of the empirical CDF.

\begin{figure}
 \epsscale{.80}
 \plotone{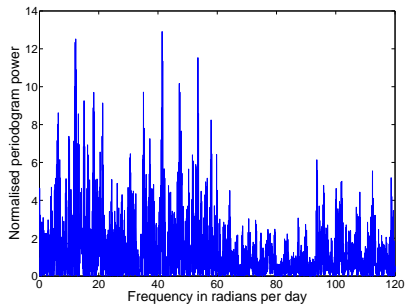}
 \caption{
Periodogram of V400 Car data after prewhitening by $f_1$ through to $f_9$.
 \label{figure4} }
\end{figure}

\begin{figure}
 \epsscale{.80}
 \plotone{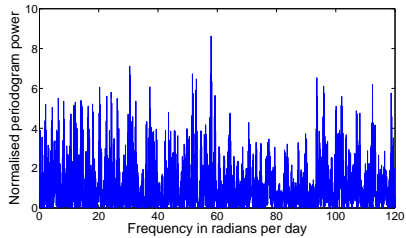}
 \caption{
Periodogram of V400 Car data after prewhitening by $f_1$ through to $f_{13}$.
 \label{figure5} }
 \end{figure}

The day-to-day and week-to-week spacings in the observations of V400 Car generate substantial
peaks in the low-frequency part of the periodogram. Once the strongest five peaks (determined
through the prewhitening procedure outlined above) are removed, the next four strongest
periodicities in the data correspond roughly to periods of three days, five days, two days and
nine days respectively and are listed as frequencies $f_6$ to $f_9$. These are probably artifacts
of the data spacing rather than real signals present in the varying luminosity of V400 Car.
However, given that the maximum possible rotation period of this star is roughly 8 days (based on
spectral line broadening measurements by \citet{balona75} and radius calculations by
\citet{engelbrecht86}), one or more of $f_6$, $f_7$, $f_8$ and $f_9$ might be related to the
rotation period of V400 Car. There are three further frequencies above a significance level of
99.7\% that could be interpreted as real periodicities in the luminosity of V400 Car. One of
these, $f_{10}$, probably corresponds to the mode reported in \citet{heynderickx94}
with a frequency of 6.61 cycles $d^{-1}$. The other, $f_{12}$, lies comfortably in the range
expected for pulsation modes in Beta Cephei stars \citep[see, for example,][]{pamyatnykh03}. We
propose that $f_{12}$ is a newly identified pulsation mode in V400 Car, bringing the total number
of detected modes in this star to seven. \citet{heynderickx94} suggested that g-modes could be
excited in Beta Cephei stars. Model calculations by \citet{pamyatnykh99} predict the existence of
pulsationally unstable low-order g-modes with eigenfrequencies between approximately 2 and 3
cycles per day for stars within the narrow range of effective temperatures covered by the four
stars discussed here. Frequencies $f_{11}$ and $f_{13}$ in Table \ref{table2} could therefore
possibly correspond to g-modes, although the significance of the latter frequency lies below the
3-sigma confidence level. If $f_{11}$ does indeed correspond to a g-mode, there will be eight
detected pulsation modes in V400 Car. The appearance of the periodogram, after prewhitening with 9
and 13 frequencies respectively, is displayed in Figures \ref{figure4} and \ref{figure5}.

\section{Significances of periodicities in V401 Car}

\begin{figure}
 \epsscale{.80}
 \plotone{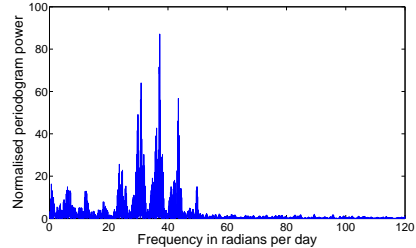}
 \caption{
Over-sampled Lomb-Scargle periodogram for V401 Car.}
 \label{figure6}
\end{figure}

\begin{figure}
 \epsscale{.80}
 \plotone{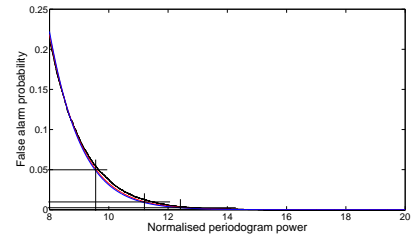}
 \caption{
Enlargement of the critical region of the CDF for V401 Car.}
 \label{figure7}
\end{figure}

The data set consists of 481 observations of V401 Car (star no. 10 in the
\citet{feast58} and \citet{turner80} designation for NGC3293) in the Johnson B
band, obtained with the 1.0 m telescope at the Sutherland station of the SAAO
during a total timespan of 65 days in 1984. The Lomb-Scargle periodogram of
the data, normalised by the variance of the data, is shown in Figure
\ref{figure6}.
The periodogram was over-sampled by a factor of approximately 25. The limiting
empirical CDF computed for the time data is very similar to the CDF for V400
Car, shown in Figure \ref{figure2}. An enlargement of the critical region of
the CDF for V401 Car is shown in Figure \ref{figure7}.
The horizontal solid lines have the same meaning as in Figure
\ref{figure3}. The power levels corresponding to the three specified
significance thresholds are displayed in Table \ref{table3}, in the
same fashion as Table \ref{table1}.
\begin{table*}
 \centering
 \begin{minipage}{140mm}
\caption{Significance Levels for V401 Car} \vspace*{4mm}
      \begin{tabular}{@{}cccc@{}}
    \hline
 Significance & \multicolumn{3}{c}{Periodogram Power Level}  \\
 & empirical & exponential function & beta function \\
   \hline
 95 \% & 9.65 & 9.57 & 9.53\\
 99 \% & 11.48 & 11.20 & 11.09\\
 99.7 \% & 12.63 & 12.41 & 12.24\\
 \hline
\end{tabular}
\label{table3}
\end{minipage}
\end{table*}
The significances of peaks appearing in the periodogram of V401 Car
were again estimated by using the exponential function providing the
best fit to the tail of the associated empirical CDF. The
oversampled, normalised Lomb-Scargle periodogram shown in Figure
\ref{figure6} was subjected to the prewhitening procedure described
in the previous section. The three observing blocks in the 65-day
observing season were zeroed to a common mean. A full list of the
results of the prewhitening procedure for V401 Car appears in Table
\ref{table4}.
\begin{table*}
 \centering
 \begin{minipage}{140mm}
\caption{Frequencies determined for V401 Car} \vspace*{4mm}
      \begin{tabular}{@{}cccccc@{}}
    \hline
 Number & Rad $d^{-1}$ & Cycles $d^{-1}$ & Amp (mmag) & Power & Significance \\
   \hline
 $f_1$ & 37.22 & 5.92 & 9.9 & 87.1 & $>99.9999$ \% \\
 $f_2$ & 35.69 & 5.68 & 6.8 & 63.2 & $>99.9999$ \% \\
 $f_3$ & 29.89 & 4.76 & 4.7 & 36.8 & $>99.9999$ \% \\
 $f_4$ & 32.07 & 5.10 & 3.9 & 34.2 & $>99.9999$ \% \\
 $f_5$ & 12.59 & 2.00 & 3.1 & 23.5 & $>99.9999$ \% \\
 $f_6$ & 0.87 & 0.14 & 2.8 & 26.6 &  $>99.9999$ \% \\
 $f_7$ & 34.30 & 5.46 & 3.0 & 27.5 & $>99.9999$ \% \\
 $f_8$ & 1.52 & 0.24 & 2.3 & 17.0 & $99.9969$ \% \\
 $f_9$ & 3.70 & 0.59 & 2.3 & 14.0 & $99.939$ \% \\
 $f_{10}$ & 31.45 & 5.01 & 2.1 & 14.7 & $99.970$ \% \\
 $f_{11}$ & 24.21 & 3.85 & 2.0 & 13.6 & $99.909$ \% \\
 $f_{12}$ & 1.14 & 0.18 & 1.7 & 11.7 & $99.39$ \% \\
 $f_{13}$ & 41.46 & 6.60 & 1.6 & 11.2 & $99.00$ \% \\
\hline
\end{tabular}
\label{table4}
\end{minipage}
\end{table*}
For convenience, each frequency is listed in units of radians per day and of
cycles per day, along with its best-fit amplitude in millimagnitudes, its
respective normalised power in the periodogram, and the associated
significance of the normalised power level, as determined from the fitting of
the exponential distribution function to the tail of the empirical CDF.

\begin{figure}
 \epsscale{.80}
 \plotone{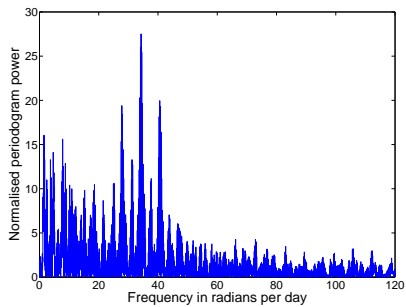}
 \caption{
Periodogram of V401 Car data after prewhitening by $f_1$ through to $f_6$.}
 \label{figure8}
\end{figure}

\begin{figure}
 \epsscale{.80}
 \plotone{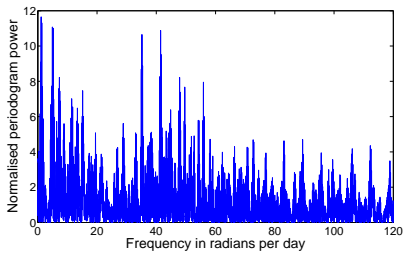}
 \caption{
Periodogram of V401 Car data after prewhitening by $f_1$ through to $f_{11}$.}
 \label{figure9}
\end{figure}

As pointed out in the previous section on V400 Car, the day-to-day and week-to-week spacings in
the observations of V401 Car generate substantial peaks in the low-frequency part of the
periodogram. Once the strongest four peaks (determined through the prewhitening procedure outlined
above) have been removed, the next two strongest periodicities in the data are a half-day period
and a seven-day period respectively. These are listed as frequencies $f_5$ and $f_6$, but are
probably artifacts due to the data spacing rather than real signals present in the varying
luminosity of V401 Car. However, $f_5$ might also be associated with a low-order g-mode, as
explained in the previous section. Frequencies $f_8$ and $f_9$ (corresponding to periods of 4 days
and 1.6 days respectively) are also likely to be artifacts due to the data spacing. However, since
the maximum possible rotation period of this star is roughly 3 days (based on spectral line
broadening measurements by \citet{balona75} and radius calculations by \citet{engelbrecht86}),
either $f_5$ or $f_9$ might be related to the rotation period of V401 Car. Aside from $f_6$, $f_8$
and $f_9$, we are left with eight frequencies above a significance level of 99.7\% that could be
interpreted as periodicities related to pulsation modes in V401 Car. Whereas the first six of
these frequencies have significances beyond the 4-sigma level (99.993\%), the remaining two (5.01
and 3.85 cycles per day respectively) are sufficiently significant that they might also be taken
into account in attempts to perform asteroseismology on V401 Car. Frequency $f_{13}$ lies well
below the 3-sigma level at a significance of 98.9\%, but its value of 6.60 cycles per day places
it well inside the expected range of excited pulsation frequencies. More intensive observation of
V401 Car might confirm this as another excited mode. The appearance of the periodogram after
prewhitening with 6 and 11 frequencies respectively, is displayed in Figures \ref{figure8} and
\ref{figure9}.

\section{Significances of periodicities in V403 Car}

\begin{figure}
 \epsscale{.80}
 \plotone{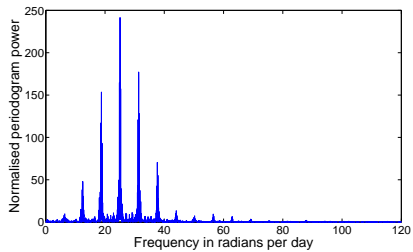}
 \caption{
Over-sampled Lomb-Scargle periodogram for V403 Car.}
 \label{figure10}
\end{figure}

\begin{figure}
 \epsscale{.80}
 \plotone{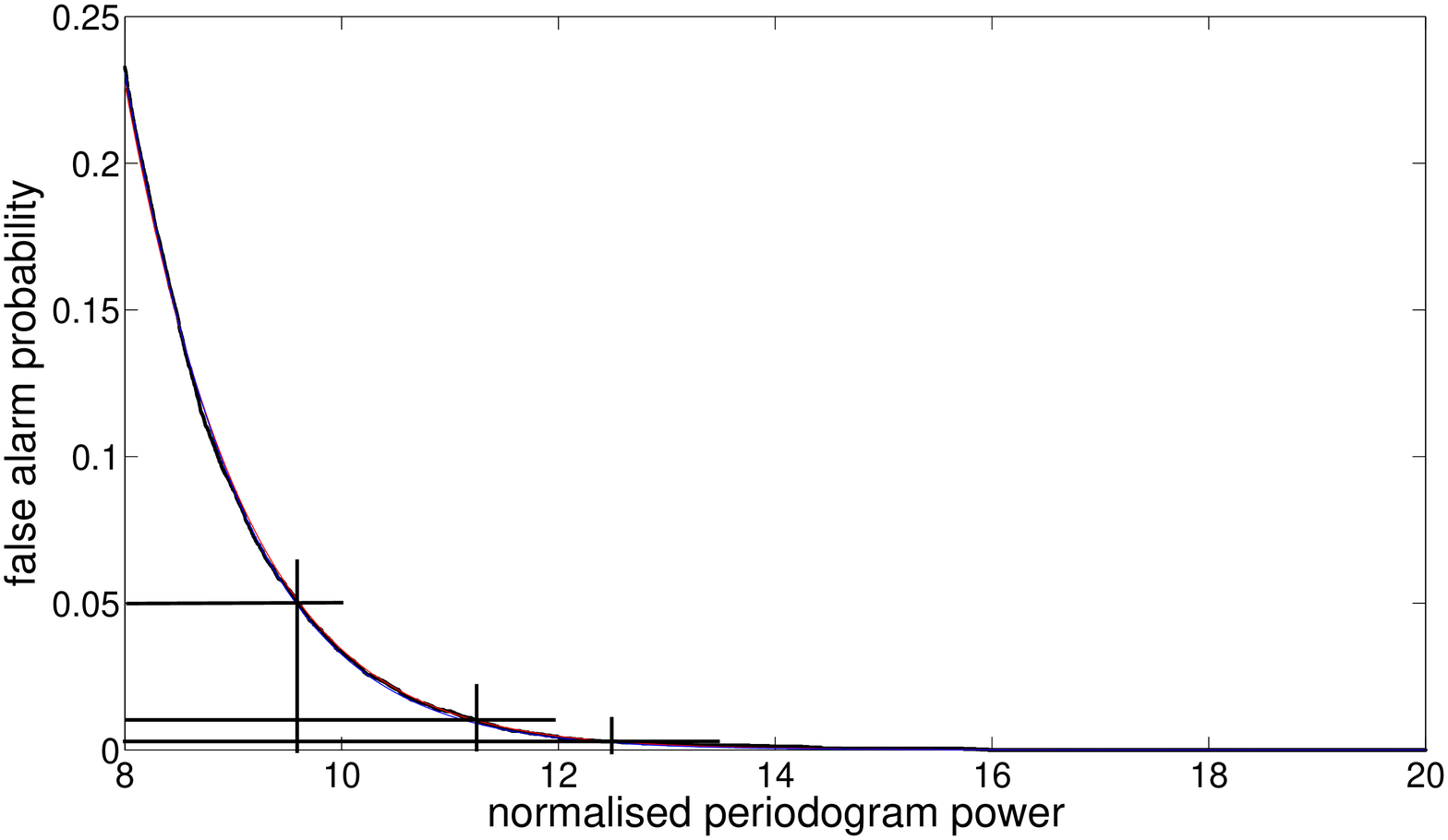}
 \caption{
Enlargement of critical region of the CDF for V403 Car.}
 \label{figure11}
\end{figure}

The data set consists of 530 observations of V403 Car (star no. 16 in the
\citet{feast58} and \citet{turner80} designation for NGC3293) in the Johnson B
band, obtained with the 1.0 m telescope at the Sutherland station of the SAAO
during a total timespan of 68 days in 1984. The Lomb-Scargle periodogram of
the data, normalised by the variance of the data, is shown in Figure
\ref{figure10}.
The periodogram was over-sampled by a factor of approximately 25. The limiting
empirical CDF computed for the time data is very similar to that of V400 Car
shown in Figure \ref{figure2}. An enlargement of the critical region of the
CDF for V403 Car is shown in Figure \ref{figure11}.
Horizontal solid lines are indicated in the same fashion as was done in Figure
\ref{figure3}. The power levels corresponding to the three specified
significance thresholds are displayed in Table 5.
\begin{table*}
 \centering
 \begin{minipage}{140mm}
\caption{Significance Levels for V403 Car} \vspace*{4mm}
      \begin{tabular}{@{}cccc@{}}
    \hline
 Significance & \multicolumn{3}{c}{Periodogram Power Level}  \\
 & empirical & exponential function & beta function \\
   \hline
 95 \% & 9.60 & 9.61 & 9.58\\
 99 \% & 11.27 & 11.24 & 11.14\\
 99.7 \% & 12.37 & 12.45 & 12.30\\
 \hline
\end{tabular}
\label{table5}
\end{minipage}
\end{table*}
The slight differences between these significances and those
appearing in \citet{frescura08} arise from the larger number of
Monte Carlo trials that produced Figure \ref{figure11} shown here,
compared to the number of trials applied in that paper.

The significances of peaks appearing in the periodogram of V403 Car
were again estimated by using the exponential function providing the
best fit to the tail of the associated empirical CDF. The
oversampled, normalised Lomb-Scargle periodogram shown in Figure
\ref{figure10} was subjected to the prewhitening procedure described
previously in this paper. The three observing blocks in the 68-day
observing season were zeroed to a common mean. A full list of the
results of the prewhitening procedure for V403 Car appears in Table
\ref{table6}.
\begin{table*}
 \centering
 \begin{minipage}{140mm}
\caption{Frequencies determined for V403 Car} \vspace*{4mm}
      \begin{tabular}{@{}cccccc@{}}
    \hline
 Number & Rad $d^{-1}$ & Cycles $d^{-1}$ & Amp (mmag) & Power & Significance \\
   \hline
 $f_1$ & 25.07 & 3.99 & 26.7 & 241.6 & $>99.9999$ \% \\
 $f_2$ & 2.57 & 0.41 & 2.9 & 31.1 & $>99.9999$ \% \\
 $f_3$ & 1.37 & 0.22 & 2.5 & 29.6 & $>99.9999$ \% \\
 $f_4$ & 30.94 & 4.92 & 2.6 & 28.2 & $>99.9999$ \% \\
 $f_5$ & 0.86 & 0.14 & 1.8 & 21.6 & $>99.9999$ \% \\
 $f_6$ & 25.74 & 4.10 & 2.0 & 20.4 &  $99.9999$ \% \\
 $f_7$ & 8.92 & 1.42 & 1.7 & 16.6 & $99.9953$ \% \\
 $f_8$ & 28.91 & 4.60 & 1.5 & 13.1 & $99.843$ \% \\
 $f_9$ & 12.32 & 1.96 & 1.2 & 10.1 & $96.9$ \% \\
 $f_{10}$ & 23.45 & 3.73 & 1.1 & 10.1 & $96.9$ \% \\

\hline
\end{tabular}
 \label{table6}
\end{minipage}
\end{table*}
For convenience, each frequency is listed in units of radians per day and of
cycles per day, along with its best-fit amplitude in millimagnitudes, its
respective normalised power in the periodogram, and the associated
significance of the normalised power level, as determined from the fitting of
the exponential distribution function to the tail of the empirical CDF. The
slight difference between these results and those appearing in
\citet{frescura08} are due to an offset of 0.009 mag in the mean of the third
block of data for this star (compared to the remainder of the data). This
offset was not corrected in the analysis appearing in \citet{frescura08}.

\begin{figure}
 \epsscale{.80}
 \plotone{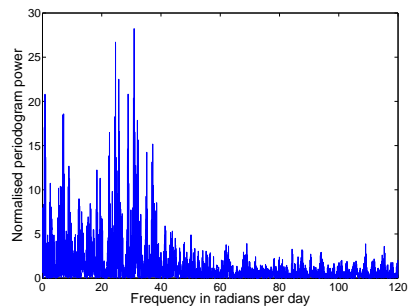}
 \caption{
Periodogram of V403 Car data after prewhitening by $f_1$ through to $f_3$.}
 \label{figure12}
\end{figure}

\begin{figure}
 \epsscale{.80}
 \plotone{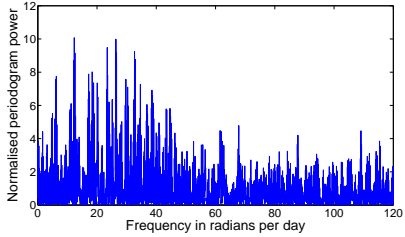}
 \caption{
Periodogram of V403 Car data after prewhitening by $f_1$ through to $f_8$.}
 \label{figure13}
\end{figure}

As pointed out in the previous two sections, the day-to-day and week-to-week spacings in the
observations of V403 Car generate substantial peaks in the low-frequency part of the periodogram.
Frequencies $f_2$, $f_3$ and $f_5$ correspond to periods of approximately 2.5 days, 4.5 days and 7
days respectively and are probably artifacts due to the data spacing, rather than real signals
present in the varying luminosity of V403 Car. Given that the maximum rotation period of this star
is roughly 10 days (based on spectral line broadening measurements by \citet{balona75} and radius
calculations by \citet{engelbrecht86}), any one of these three periodicities might be related to
the rotation period of V403 Car. Aside from $f_2$, $f_3$, $f_5$ and $f_7$, we are left with four
frequencies above a significance level of 99.7\% that could be interpreted as periodicities
related to pulsation modes in V403 Car and might be taken into account in attempts to perform
asteroseismology on this star. Frequencies $f_9$ and $f_{10}$ lie below the 3-sigma level at a
significance of 97.0\%, but their values of 1.96 cycles per day and 3.73 cycles per day place them
inside the expected range of excited pulsation frequencies for low-order g-modes and p-modes
respectively. More intensive observation of V403 Car might confirm them as excited modes. The
appearance of the periodogram after prewhitening with 3 and 8 frequencies respectively, is
displayed in Figures \ref{figure12} and \ref{figure13}.

\section{Significances of periodicities in V405 Car}

\begin{figure}
 \epsscale{.80}
 \plotone{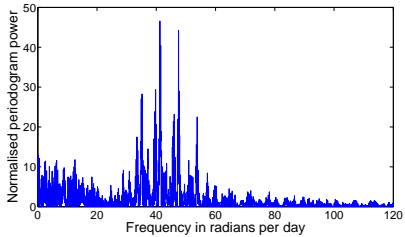}
 \caption{
Over-sampled Lomb-Scargle periodogram for V405 Car.}
 \label{figure14}
\end{figure}

\begin{figure}
 \epsscale{.80}
 \plotone{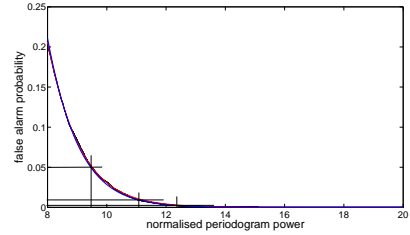}
 \caption{
Enlargement of critical region of the CDF for V405 Car.}
 \label{figure15}
\end{figure}

The data set consists of 357 observations of V405 Car (star no. 14 in the
\citet{feast58} and \citet{turner80} designation for NGC3293) in the Johnson B
band, obtained with the 0.5 m, 0.75 m and 1.0 m telescopes at the SAAO during
a total timespan of 70 days in 1983. The Lomb-Scargle periodogram of the data,
normalised by the variance of the data, is shown in Figure \ref{figure14}.
The periodogram was over-sampled by a factor of approximately 25. The limiting
empirical CDF computed for the time data is very similar to that of V400 Car
shown in Figure \ref{figure2}. An enlargement of the critical region of the
CDF for V405 Car is shown in Figure \ref{figure15}.
Horizontal solid lines are indicated in the same fashion as was done in Figure
\ref{figure3}. The power levels corresponding to the three specified
significance thresholds are displayed in Table \ref{table7}.
\begin{table*}
 \centering
 \begin{minipage}{140mm}
\caption{Significance Levels for V405 Car} \vspace*{4mm}
      \begin{tabular}{@{}cccc@{}}
    \hline
 Significance & \multicolumn{3}{c}{Periodogram Power Level}  \\
 & empirical & exponential function & beta function \\
   \hline
 95 \% & 9.49 & 9.51 & 9.46\\
 99 \% & 11.02 & 11.14 & 10.99\\
 99.7 \% & 12.11 & 12.34 & 12.12\\
 \hline
\end{tabular}
\label{table7}
\end{minipage}
\end{table*}

The significances of peaks appearing in the periodogram of V405 Car
were again estimated by using the exponential function providing the
best fit to the tail of the associated empirical CDF. The
oversampled, normalised Lomb-Scargle periodogram shown in Figure
\ref{figure14} was subjected to the prewhitening procedure described
previously in this paper. The three observing blocks in the 70-day
observing season were zeroed to a common mean. A full list of the
results of the prewhitening procedure for V405 Car appears in Table
\ref{table8}.
\begin{table*}
 \centering
 \begin{minipage}{140mm}
\caption{Frequencies determined for V405 Car} \vspace*{4mm}
      \begin{tabular}{@{}cccccc@{}}
    \hline
 Number & Rad $d^{-1}$ & Cycles $d^{-1}$ & Amp (mmag) & Power & Significance \\
   \hline
 $f_1$ & 41.22 & 6.56 & 5.0 & 46.6 & $>99.9999$ \% \\
 $f_2$ & 39.76 & 6.33 & 3.6 & 31.7 & $>99.9999$ \% \\
 $f_3$ & 37.09 & 5.90 & 2.8 & 22.8 & $>99.9999$ \% \\
 $f_4$ & 0.16 & 0.03 & 3.7 & 26.8 & $>99.9999$ \% \\
 $f_5$ & 2.36 & 0.38 & 2.3 & 19.1 & 99.9997 \% \\
 $f_6$ & 50.94 & 8.11 & 1.8 & 14.3 & 99.958 \% \\

 \hline
\end{tabular}
\label{table8}
\end{minipage}
\end{table*}
For convenience, each frequency is listed in units of radians per day and of
cycles per day, along with its best-fit amplitude, its respective
normalised power in the periodogram, and the associated significance of that
power level, as determined from the fitting of the exponential distribution
function to the tail of the empirical CDF.

\begin{figure}
 \epsscale{.80}
 \plotone{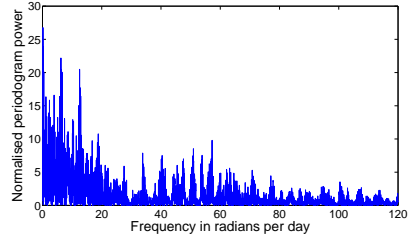}
 \caption{
Periodogram of V405 Car data after prewhitening by $f_1$ through to $f_3$.}
 \label{figure16}
\end{figure}

\begin{figure}
\epsscale{.80} \plotone{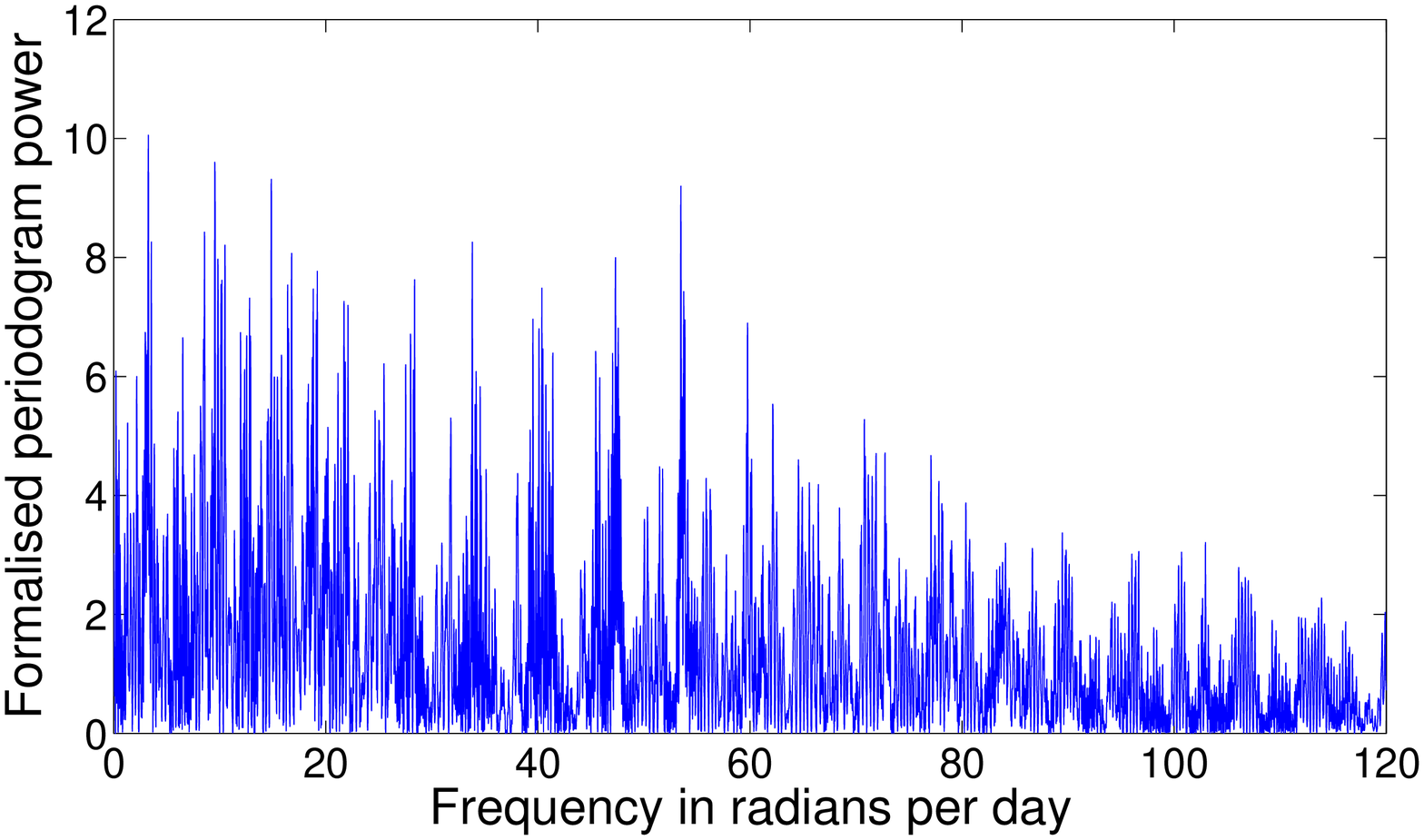}
 \caption{
Periodogram of V405 Car data after prewhitening by $f_1$ through to $f_6$.}
 \label{figure17}
\end{figure}

The appearance of the periodogram after prewhitening with 3 and 6
frequencies respectively, is displayed in Figures \ref{figure16} and
\ref{figure17}.

As pointed out in the previous sections, the day-to-day and week-to-week spacings in the
observations of V405 Car generate substantial peaks in the low-frequency part of the periodogram.
Frequencies $f_4$ and $f_5$ correspond to periods of approximately 40 days and 3 days respectively
and are probably artifacts due to the data spacing rather than to real signals present in the
varying luminosity of V405 Car, although $f_5$ might be associated with the rotation period of
V405 Car, which has an upper limit of roughly 3 days (based on spectral line broadening
measurements by \citet{balona75} and radius calculations by \citet{engelbrecht86}). Apart from
$f_4$ and $f_5$, we are left with four frequencies above a significance level of 99.7\% that might
be interpreted as periodicities related to pulsation modes in V405 Car and which could be taken
into account in attempts to perform asteroseismology on this star. Frequency $f_6$ has not been
reported in the literature before.

\section{Discussion}

We have used the significance test described in \citet{frescura08} to identify
six previously unreported periodicities which might correspond to pulsation
modes in the Beta Cephei stars V400 Car, V401 Car, V403 Car and V405 Car. The
new identifications are listed in Table \ref{table9}.
\begin{table*}
 \centering
 \begin{minipage}{140mm}
\caption{ Potential new pulsation modes identified in V400 Car, V401 Car, V403 Car and V405 Car}
\vspace*{4mm}
      \begin{tabular}{@{}cccc@{}}
    \hline
 Star & Frequency in Rad $d^{-1}$ & Frequency in Cycles $d^{-1}$ & Amplitude (mmag)\\
   \hline
V400 Car & 53.48 & 8.51 & 1.1 \\
V401 Car & 12.59 & 2.00 & 3.1 \\
V401 Car & 31.45 & 5.01 & 2.1 \\
V401 Car & 24.21 & 3.85 & 2.0 \\
V403 Car & 28.91 & 4.60 & 1.5 \\
V405 Car & 50.94 & 8.11 & 1.8 \\
\hline
\end{tabular}
\label{table9}
\end{minipage}
\end{table*}

The periodograms shown in Figures \ref{figure1}, \ref{figure6}, \ref{figure10} and \ref{figure14}
all appear to display a rising background noise level below frequencies of about 20 radians per
day. This may seem to imply that our data contains what is traditionally described as ``red
noise". However, after removal of all significant low frequency components in the prewhitening
procedure, the periodograms display a behaviour that is more typical of ``white noise" (not shown
above). We take this to be an indication that the apparent redness of the original data is an
artefact of the presence of a few strong low frequency components rather than a dense noise
spectrum, and that the correct noise model to use in the corresponding Monte Carlo simulations is
that of white rather than red noise. In any event, our principal interest is in potential
pulsation modes, and the data clearly display ``flat" noise levels in the frequency ranges
predicted for low-order p-modes. Thus the significances listed in Tables \ref{table2},
\ref{table4}, \ref{table6} and \ref{table8} (at least for the p-modes) are very unlikely to be far
off the values that would be obtained with a more refined noise model for the Monte Carlo
simulations.

We conclude by stating that our data allow a 3-sigma detection of seven probable pulsation modes
in V400 Car, another eight in V401 Car, four modes in V403 Car and four modes in V405 Car. One of
these modes, $f_5$ in V401 Car, is associated with a frequency value that lies in the expected
range of eigenfrequencies of low-order g-modes. This corresponds with suggestions made by
\citet{heynderickx94} and with the results of model calculations by \citet{dziembowski93} and
\citet{pamyatnykh99,pamyatnykh03} respectively. Further low-amplitude modes in these stars might
well remain to be discovered amongst the noise levels present in our data. The open cluster
NGC3293 continues to present itself as a rich source of data with which models of Beta Cephei
pulsation may be tested. Continued, intensive observation of the Beta Cephei stars in this cluster
appears imperative.

\acknowledgments

We thank Michel Breger and Gerald Handler for useful discussions. We also thank the South African
SKA Office in Johannesburg for use of their facilities.

\end{document}